%% file: main.tex
\documentclass[12pt]{article}

\input{packages}

\usepackage{graphics}
\usepackage{amsthm}
\usepackage{amssymb}
\usepackage{amsmath}
\usepackage{amsfonts}
\usepackage{epic}
\usepackage{hyperref}
\usepackage{epsfig}
\usepackage{bm}
\usepackage{amscd}

\theoremstyle{plain}

\theoremstyle{definition}

\def\be{\begin{equation}}
\def\ee{\end{equation}}
\smallskip

\begin{document}

\headsep=-0.5cm

\begin{titlepage}
\begin{flushright}
%hep-th/.......\\
\end{flushright}
%%%%%%%%%%%%%%%%%%%%%%%%%%%%%%%%%%%%%%%%%%%%%%%%%%%%%%%%%%%%%%%%%%%%%%%%
\begin{center}
\noindent{{\LARGE{Probing the near-horizon geometry of black rings}}}
%\noindent{{\LARGE{Near the black ring horizon}}}

\smallskip
\smallskip
\smallskip

\smallskip
\smallskip

\smallskip

\smallskip

\smallskip
\smallskip
\noindent{\large{Gaston Giribet$^{1}$, Juan Laurnagaray$^{2}$, Pedro Schmied$^{2}$}}
\end{center}

\smallskip

\centerline{$^1$ Department of Physics, New York University, NYU}
\centerline{{\it 726 Broadway, New York, NY10003, USA.}}

\smallskip
\smallskip

\centerline{$^2$ Department of Physics, University of Buenos Aires, UBA and IFIBA, CONICET}
\centerline{{\it Ciudad Universitaria, pabell\'on 1, 1428, Buenos Aires, Argentina.}}

\smallskip
\smallskip

\smallskip
\smallskip

\smallskip

\smallskip
\begin{abstract}
In many cases, the near-horizon geometry encodes sufficient information to compute conserved charges of a gravitational solution, including thermodynamic quantities. These charges are Noether charges associated to asymptotic isometries that preserve appropriate boundary conditions at the future horizon. For isolated, compact horizons these charges turn out to be integrable, conserved and finite, and they have been studied in many examples of interest, notably in 3+1 dimensions. In higher dimensions, where the variety of horizon structures is more diverse, it is still possible to apply the same method, although explicit examples have so far been limited to simple topologies. In this paper, we demonstrate that such computations can also be applied to higher-dimensional solutions with event horizons whose spacelike cross sections exhibit non-trivial topology. We provide several explicit examples, with particular focus on the 5-dimensional black ring.
\end{abstract}

\end{titlepage}
%%%%%%%%%%%%%%%%%%%%%%%%%%%%%%%%%%%%%%%%%%%%%%%%%%%%%%%%%%%%%%%%%%%%%%%%

\newpage

%\documentclass[12pt, a4paper]{article}

%\input{packages}

%\title{Black Ring}
%\author{Juan Laurnagaray, Pedro Schmied}
%\date{\today}

%\begin{document}

%\maketitle

 \section{Introduction}

The black hole topology theorem, as originally presented by Hawking in \cite{Hawking:1971vc}, states that the cross section of the event horizon of a stationary black hole in 4-dimensional spacetime, provided the dominant energy condition is satisfied, must have topologically spherical boundary. More precisely, each connected component of the 2-surface in the event horizon has to have topology $S^2$. This theorem has been a cornerstone of black hole theory for decades, providing a powerful constraint on the possible shapes that black holes can take in the quasi-stationary regime of general relativity. The topology theorem is motivated by the physical requirement that outgoing null geodesics from the horizon must not converge in the future, as this would imply the existence of a caustic in the spacetime: If the horizon had any other topology besides that of a sphere, it would be possible to find an outer-trapped 2-surface outside the horizon; that is to say, it would be possible to deform the horizon outwards into the exterior region in such a way that the future-directed outgoing null geodesics orthogonal to it would be converging, this leading to an inconsistency. This seems to be a quite robust theorem that has been exhaustively revised and generalized \cite{Browdy:1995qu, Gannon:1976, Jacobson:1994hs, Chrusciel:1994tr}; however, it is only valid in dimension $4$. In spacetime dimension $5$, the topology censorship argument does not hold and, in addition to the Myers-Perry solution, which describes a family of stationary axisymmetric black holes with horizon topology of $S^3$, solutions with event horizons of different topology also exist \cite{Galloway:2005mf}. The first explicit example of a smooth, non-spherical black hole solutions was discovered by Emparan and Reall in 2001 \cite{Emparan:2001wn}; this is the black ring. The black ring has a horizon topology of $S^1 \times S^2$, which is fundamentally different from that of the Myers-Perry black holes and is a result of the non-trivial topology of the extra dimension. The discovery of black rings was a significant breakthrough in the study of black holes in higher dimensions and has led to numerous subsequent studies and investigations into their properties and behavior \cite{Emparan:2008eg, Emparan:2006mm, Emparan:2004wy} as well as their generalizations \cite{Elvang:2004rt, Elvang:2007rd}. In this paper, we are interested in investigating the geometry of the black ring from a different viewpoint: we will study the symmetries that the black ring exhibit in their near horizon region and the associated conserved charges. 

In \cite{Hawking:2015qqa}, Hawking proposed that the geometry describing the near horizon region of black holes must exhibits an infinite-dimensional symmetry known as supertranslation. This was made precise in \cite{Donnay:2015abr}, where it was shown that, indeed, black holes in 4-dimensional spacetime do exhibit infinite-dimensional symmetry near their horizons. By prescribing a physically sensible set of boundary conditions at the horizon, the authors of \cite{Donnay:2016ejv} derived the algebra of asymptotic Killing vectors, which turns out to be infinite-dimensional: it includes the supertranslations conjectured by Hawking together with the 2-dimensional local conformal algebra. The Noether charges associated to the asymptotic diffeomorphisms that preserve the boundary conditions at the (isolated) horizon are integrable, finite and conserved, and they carry important information about the black hole, like the angular momentum, gauge charges, and entropy. The computation of horizon charges done in \cite{Donnay:2016ejv} was later generalized to higher dimensions in \cite{Shi:2016jtn}, although explicit examples have so far been limited to simple topologies. Here, we will demonstrate that such computations can also be applied to solutions with non-trivial topology, such as topological black holes, black strings and branes, and the black ring.

The paper is organized as follows: In section 2, in order to prepare the ground, we address a simple example of black hole solution with non-trivial topology: we compute the near horizon charges of topological black holes in asymptotically AdS$_5$ spacetime, showing that one of them, the one associated to global translations in the advanced time at the horizon, correctly reproduces the Bekenstein-Hawking entropy. In section 3, we briefly review the asymptotic boundary conditions and the associated Noether charges for an arbitrary 5-dimensional black object. In section 4, we study the near-horizon geometry of the black ring solutions, including the cases with angular momentum along the $S^1$ direction and along one of the directions of the $S^2$. We explicit find the coordinate system to accommodate the black ring geometry in the asymptotic conditions defined in \cite{Donnay:2016ejv}, we discuss the symmetries and we compute the charges at the horizon. We also perform the analysis for static configurations that exhibit horizons with defects.

\section{Topological black holes}

The simplest examples of a black hole solution with horizon of non-trivial topology is the often-called topological black hole \cite{Birmingham:1998nr}. A particular case with locally flat horizon is given by
\begin{equation}
    ds^2 = - H(r)\, dv^2+2\, dvdr+r^2\delta_{AB}\, dx^Adx^B \label{LaUno}
\end{equation}
with $t\in \mathbb{R}$, $r\in \mathbb{R}_{\geq 0}$, and $x^A\sim x^A+2\pi \lambda_A$, $\lambda_A \in \mathbb{R}_{>0}$, $A,B=1,2,3$. $\delta_{AB}$ is the Kronecker tensor, so that the topology of the constant-$v$ slices of the horizon is $S^1\times S^1\times S^1$. The function $H(r)$ is given by 
\begin{equation}
    H(r)= \frac{r^4-r^{4}_+}{L^2 r^{2}} \, ,\label{LaDos}
\end{equation}
where $r_+\in \mathbb{R}_{\geq 0 }$ is an integration constant that corresponds to the location of the event horizon location.  $L $ is a dimension $-1$ parameter related to the cosmological constant $\Lambda =-6/L^2$. Metric (\ref{LaUno})-(\ref{LaDos}) is a solution to Einstein equations with negative cosmological constant that describes asymptotically locally AdS$_5$ black holes written in Eddington-Filkenstein type coordinates. The mass of the solution is given by
\begin{equation}
    M=\frac{3\pi^2r_+^4}{2GL^2}\, \lambda_1\lambda_2\lambda_3 \, .
\end{equation}

The form that the metric takes in the near horizon limit is easily obtained by defining the coordinate $\rho =r-r_+$ and considering the small $\rho $ limit. This yields
\begin{equation}
    ds^2 \simeq - \frac{4r_+}{L^2 }\rho \, dv^2+2\, dv \, d\rho +r^2_+\delta_{AB\, }dx^Adx^B + ...
\end{equation}
where the ellipsis stand for subleading orders in the the near horizon expansion $\rho \simeq 0$. A straightforward application of the formalism of \cite{Donnay:2016ejv} to this particular case yields the following result for the Noether charge
\begin{equation}
    Q[\partial_v ]\, =\, \frac{1}{16\pi G}\lim_{\rho \to 0}\int_{H} d^3 x\, \sqrt{\det g_{AB}}\,\rho^{-1}g_{vv}\, =\, \frac{2\pi^2 r_+^4}{L^2 G } \,  \lambda_1\lambda_2\lambda_3 \, . \label{Lacarguita}
\end{equation}
This is the Noether charge associated to the rigid translations in the retarded time $v$ on the horizon. The integral is over the constant-$v$, constant-$\rho$ sections of the horizon in the limit $r\to r_+$, and this is what the subindex $H$ refers to. Charge (\ref{Lacarguita}) coincides with Wald entropy \cite{Wald:1993nt} and so it gives the product of the Hawking temperature and the Bekenstein-Hawking entropy of the topological black hole; namely
\begin{equation}
    T=\frac{\kappa }{2\pi }=\frac{r_+}{\pi L^2 }
\ \ , \ \  \ \ \ \ 
    S=\frac{\text{A}}{4G}=\frac{2\pi^3r_+^ 3\lambda_1\lambda_2\lambda_3}{G}\, , 
\end{equation}
respectively. $\kappa $ is the surface gravity and $\text{A}$ is the volume of the 3-torus at the horizon ($\rho = 0, \, v=\text{const}$). Entropy (\ref{Lacarguita}) is a particular case of the infinite set of conserved charges defined  in \cite{Donnay:2015abr, Donnay:2016ejv}. The calculation can easily be extended to other 5-dimensional solutions, such as black holes with horizons of negative constant curvature. The latter are constructed by shifting $H(r)\to H(r)-1$ in (\ref{LaDos}) and replacing the Kronecker symbol in the component $g_{AB}$ in (\ref{LaUno}) by a constant curvature metric on the quotient space $\mathbb{H}_3/\Gamma$, being $\Gamma $ a discrete subgroup of the isometries of the unit hyperbolic space $\mathbb{H}_3$. For those quotients yielding non-compact horizons, the solutions can be regarded as black 3-branes, for which the computation of charges and thermodynamic quantities per unit of volume is still well defined. The case of Myers-Perry black holes with two independent angular momenta can also be addressed \cite{Shi:2016jtn}. Here, we are interested in the black ring, which is notably more involved: For the solutions with horizons of topology $S^1\times S^2$, we will show that: $a)$ the asymptotic form of the metric is compatible with the near-horizon analysis done in \cite{Donnay:2015abr, Donnay:2016ejv}, which implies that the black ring exhibits an infinite symmetry enhancement in its vicinity; $b)$ the zero modes of Noether charges associated to such symmetries correctly reproduce the conserved quantities and thermodynamic variables of the black ring.

\section{Symmetries and Noether charges}

In this section we briefly review the asymptotic boundary conditions and the associated Noether charges in the 5-dimensional case. We do this in order to introduce our notation; the details can be found in \cite{Donnay:2015abr, Donnay:2016ejv, Shi:2016jtn}. 

Our goal is to study the black ring geometry in the near horizon limit. In order to do so, we need to find a system of coordinates that, near the horizon, makes it evident that the geometry satisfies the boundary conditions prescribed in \cite{Donnay:2016ejv}. This would imply that in the vicinity of the black ring the asymptotic Killing vectors form an infinite-dimensional algebra. 

Near the horizon, consider the following expansion in powers of $\rho $
\begin{equation}\label{Desarrollo}
    g_{\mu \nu} = g_{\mu \nu}^{(0)}\, +\, g_{\mu \nu}^{(1)} \, \rho \, +\,  g_{\mu \nu}^{(2)} \, \rho^2 \, +\, \mathcal O(\rho^3)
\end{equation}
where $g_{\mu \nu}^{(n)}$ are functions of $v$ and $x^A$, $A=1,2,3$, and are independent of $\rho $. The horizon is at $\rho = 0$. Next, we impose boundary conditions defined by
\begin{equation}
g^{(0)}_{vv}=0\, , \ \ \ g^{(1)}_{vv}=-2\kappa \, , \ \ \  g^{(0)}_{vA}=0, 
\end{equation}
together with the gauge condition 
\begin{equation}\label{Desarrollo2}
g_{\rho v}=-1 \, , \ \ \   g_{\rho \rho}=0 \, , \ \ \   g_{\rho A}=0\, ; 
\end{equation}
constant $\kappa $ is the surface gravity. Boundary conditions (\ref{Desarrollo})-(\ref{Desarrollo2}) are the near horizon form studied in \cite{Donnay:2015abr, Donnay:2016ejv} and they were considered in many different contexts \cite{Moncrief:1983xua, Booth:2012xm}. It was shown in \cite{Donnay:2015abr, Donnay:2016ejv, Shi:2016jtn} that the asymptotic Killing vectors preserving these asymptotic boundary conditions (\ref{Desarrollo})-(\ref{Desarrollo2}), but allowing the subleading components in the small $\rho $ expansion to change, form an infinite-dimensional algebra that includes supertranslations generated by $P(x^A)\partial_{v}$ together with transformations generated by $L^A(x^A)\partial_{x^A}$, with $P$ and $L^A$ being arbitrary functions of the angular variables. These asymptotic Killing vectors have associated the following Noether charges
\begin{equation}
	Q[P, L^A] = \frac{1}{16 \pi G} \int_H d^3x\,\sqrt{\text{det}{g_{\text{AB}}^{(0)}}}\,\of{2 \kappa\,P - L^A g_{v\text{A}}^{(1)}}\,,
  \label{eq_cargas_Barnich}
\end{equation}
where, again, we are using Latin indices $A, B=1,2,3$ to denote the angular coordinates; we will later use coordinates $x^1=x$, $x^2=\phi$, and $x^3=\psi$, with $\partial_{\phi}$ and $\partial_{\psi}$ being the Killing vectors corresponding to two independent angular momenta. The charges are calculated as integrals over constant-$v$ slices on the horizon. 

For sufficiently symmetric solutions, the relevant information is encoded in the zero-modes of the charges, for which $P$ and $L^A$ are constants. In particular, we find
\begin{align}
    Q\off{1, 0} = T\,S\hspace{0.4cm},\hspace{0.4cm}Q[0, \delta^A_{\psi}]  = J_{\psi}\hspace{0.4cm},\hspace{0.4cm}Q[0, \delta^A_{\phi}]  = J_{\phi}\,.
\end{align}
where $J_{\psi}$ and $J_{\phi}$ refer to the angular momentum around $\psi $ and $\phi $, respectively. $T= \kappa / (2\pi)$ is the Hawking temperature and $S=\text{A}/(4G)$ is the Bekenstein-Hawking entropy. In the next section we will see how this can be applied to the black ring.

\section{Black Ring}

Black rings are solution to Einstein equations in dimension greater than 4 that describe black holes with an event horizon that has a ring shape. In contrast to Myers-Perry black hole, which has an event horizon with the topology of the 3-sphere, the black ring has an event horizon that is toroidal in shape, resembling a hyperdonut of topology $S^1\times S^2$. We will focus on the case in which the solution only has angular momentum along the $S^1$ direction. Later, we will consider the case in which the ring exhibits angular momentum in the $S^2$. 

\subsection{Rotation along the $S^1$ direction}

The metric of a neutral black ring rotating in the $S^1$ direction \cite{Emparan:2006mm} can be written as
\begin{align}
\label{eq:black_ring_near_horizon_coord:black_ring_cmetric_coord}
	ds^2 = \frac{R^2 F\of x}{(x - y)^2} 
	\off{- \frac{G\of y}{F\of y} d\psi^2 - \frac{dy^2}{G\of y} + \frac{dx^2}{G\of x} + \frac{G\of x}{F\of x} d\phi^2} - \frac{F\of y}{F\of x} \of{dt - C \, R \, \frac{1 + y}{F\of y} \, d \psi}^2\,,
\end{align}
where
\begin{align}
	& F\of \xi = 1 + \lambda \, \xi, & & G\of \xi = (1 - \xi^2)(1 + \nu \, \xi), & & C = \sqrt{\lambda (\lambda - \nu) \frac{1 + \lambda}{1 - \lambda}}, & &  0 < \nu \leq \lambda < 1 \, .
\end{align}
$R$, $\lambda $, and $\nu $ are constants: by taking both $\lambda$ and $\nu$ to zero, flat spacetime is recovered. This observation is useful to gain intuition about the coordinate system. The coordinates are such that the asymptotic infinity is at $x, y \to -1$. The event horizon is located where the function $G\of y$ vanishes; we will denote this root $y_H$ and hereafter we will denote $\Delta\of y \equiv G\of y$. At $x = -1$, the period of $\phi$ for the horizon to be regular is $\delta \phi = 2 \pi \sqrt{1 - \lambda}/(1 - \nu)$; on the other hand, at $x = +1$ the analogous requirement demands $\delta \phi = 2 \pi \sqrt{1 + \lambda}/(1 + \nu)$; we also need to regularize at $y=-1$. These regularity conditions are compatible if only if the following relation between parameters holds
\begin{equation}
    \lambda = \frac{2 \nu}{1 + \nu^2},
\end{equation}
yielding the periods
\begin{align}
	\label{eq:black_ring_near_horizon_coord:no_conic_sing_condition}
	 \delta\phi = \delta \psi = 2 \pi \frac{\sqrt{1 - \lambda}}{1 - \nu} = \frac{2 \pi}{\sqrt{1 + \nu^2}}.
\end{align}
That is to say, the black ring has effectively only two free parameters: $\nu$ and $R$; while the former gives the ratio between the radii of the $S^1$ and the $S^2$, the latter gives the radius of the $S^1$. The parameter $\lambda$ carries information about the rotation speed, and the fact that it is determined by $\nu$ through the regularity condition means that the solution is supported by its own rotation. This can be understood in many different ways: Notice that the coordinates in (\ref{eq:black_ring_near_horizon_coord:black_ring_cmetric_coord}) are reminiscent of the ones used to describe the $C$-metric in 4 dimensions. In fact, the black ring can be thought of as a non-trivial dimensional oxidation of a pair of accelerated black holes in the presence of electromagnetic background in dimension 4. Therefore, regularity of the black ring horizon is the 5-dimensional analog of tuning the acceleration of the 4-dimensional charged black hole to match the electromagnetic force per unit of mass.

In order to study the black ring in the near horizon limit, we need to find the explicit coordinate transformation that brings the metric into the form that obeys the boundary conditions prescribed in \cite{Donnay:2016ejv}. In order to achieve so, we follow the approach of \cite{Booth:2012xm}. First, we rewrite the metric as follows
\begin{align}
	ds^2 & = \Lambda\of{x, y} R^2 \off{\frac{ \Delta\of y}{\Sigma\of{x, y}}dt^2 - \frac{dy^2}{\Delta\of y} + \frac{dx^2}{G\of x} + \frac{G\of x}{F\of x} d\phi^2} + g_{\psi \psi} (d\psi - \omega\of{x, y} \, dt)^2
\end{align}
where we have defined the functions
\begin{align}
    \Lambda\of{x, y} & = \frac{F\of x}{(x - y)^2}\hspace{0.2cm},\hspace{0.4cm}g_{\psi \psi} = -\frac{C^2 R^2(1 + y)^2}{F\of x F\of y} - \frac{\Delta\of{y}\,R^2 \Lambda\of{x, y}}{F\of y}\,\hspace{0.2cm},\hspace{0.2cm}
    \omega\of{x, y} = - \frac{C R (1 + y)}{F(x)\,g_{\psi \psi}}\,\hspace{0.2cm}, \nonumber\\
    \Sigma\of{x, y} & = - \Lambda\of{x, y} R^2 \, \frac{C^2 (1 + y)^2 (x - y)^2 + \Delta\of y F\of x^2}{F\of x F\of y}\,.
\end{align}
It is easy to check that the functions $\Sigma\of{x, y}$ and $\omega\of{x, y}$ take constants values when evaluated on the horizon; that is, they do not depend on the variable $x$ when are evaluated at $y=y_H$. Second, we propose the following change of coordinates
\begin{equation}
    \begin{aligned}
        \psi' & = \psi - \omega_H\,t, \\
        v & = t + \int_{y_H}^y dy'\frac{\sqrt{\Sigma\of{x, y'}}}{\Delta\of{y'}} \\
        \tilde \psi & = \psi' + \int_{y_H}^y dy' \of{\omega\of{x, y'} - \omega_H} \frac{\sqrt{\Sigma\of{x, y'}}}{\Delta\of{y'}}\,.
    \end{aligned}
    \label{eq:coordinate_change}
\end{equation}
which, in differential form, read
\begin{equation}
    \begin{aligned}
        d \psi' & = d\psi - \omega_H\,dt, \\
        dv & = dt + \frac{\sqrt{\Sigma\of{x, y}}}{\Delta \of{y}} dy + \gamma\of{x, y} dx, \\
        d\tilde \psi & = d\psi' + \of{\omega\of{x, y} - \omega_H} \frac{\sqrt{\Sigma\of{x, y}}}{\Delta\of{y}} dy + h\of{x, y} dx \, ,
    \end{aligned}
    \label{eq:differential_coordinate_change}
\end{equation}
with
\begin{align}
	\gamma\of{x, y} & = \int_{y_H}^y \frac{dy'}{\Delta\of{y'}} \, \sqrt{\Sigma\of{x, y'}} \, , \nonumber \\ 
    \hspace{0.2cm} h \of{x, y} & = \int_{y_H}^y \frac{dy'}{\Delta\of{y'}} \, \partial_x\off{\of{\omega\of{x, y} - \omega_H}\sqrt{\Sigma\of{x, y'}}} \, .
\end{align}
$\omega_H$ stands for the value of the function $\omega (x,y)$ at the horizon. In these new coordinates, the black ring metric reads
\begin{align}
    ds^2  & = g_{\psi \psi} \of{d \tilde \psi - (\omega\of{x, y} - \omega_H) \, dv + \off{(\omega\of{x, y} - \omega_H) \, \gamma\of{x, y} - h\of{x, y}} \, dx}^2 \nonumber \\
    & \quad + \Lambda\of{x, y} \, R^2 \Bigg[
    \frac{\Delta\of y}{\Sigma\of{x, y}} \, dv^2 - \frac{2 \, dv \, dy}{\sqrt{\Sigma\of{x, y}}} - 2
    \frac{\gamma\of{x, y} \Delta\of y}{\Sigma\of{x, y}} \, dv \, dx
     \nonumber \\
    &  - 2 \frac{\gamma\of{x, y}}{\sqrt{\Sigma\of{x, y}}} \, dy \, dx + \frac{G(x)}{F(x)}\, d\phi^2 + \of{\frac{\Delta\of y}{\Sigma\of{x, y}} \, \gamma\of{x, y}^2 + \frac{1}{G(x)}} dx^2 \Bigg]
    \label{eq:black_ring_rotating_circle:metric:last_coordinate_change}
\end{align}
Provided the correct angular periods are chosen, this coordinate system is regular at the horizon, where the following metric is induced
\begin{align}\label{eq:S1_metrica_inducida}
ds^2_{|H}  = \Lambda_H R^2 \off{\frac{dx^2}{G(x)} + \frac{G(x)}{F(x)}\, d\phi^2} + g_{\psi \psi  |H}\,d\tilde{\psi}^2\,.
\end{align}
The subindex $H$ means that the functions are evaluated at the horizon. Then, the horizon can be thought of as a 3-dimensional spacelike hypersurface given by constant-$v$ slices with $\ell = \partial_v$ being a null vector on it. We can also find a second null vector on the horizon, normalized such as $\ell_{\mu }n^{\mu } = 1$; namely
\begin{align}
    n = - \frac{\sqrt{\Sigma_H}}{R^2 \Lambda_H}\,\partial_y \, .
\end{align}
Now, we consider a family of geodesics that cross the horizon, whose tangent vector is $n^{\mu}$ and with affine parameter $\rho$, such that $\rho_{|H} = 0$. Up to first order in $\rho$, this congruence of curves defines the vector field
\begin{equation}
    \Xi^{\mu }(v,\rho, \theta, \phi, \tilde{\psi}) = \{v, y_H,\theta,\phi, \tilde{\psi} \}^{\mu } + \rho \, n^{\mu} + \mathcal{O}(\rho^2)\,.
\end{equation}
Using this, we compute the Lie derivative of the metric along the geodesics of tangent $n^{\mu }$, so the expansion of the full metric will have the form (\ref{Desarrollo}) with $g_{\mu \nu}^{(0)} = g_{\mu \nu\, |H}$ and $g_{\mu \nu}^{(1)} = (\mathcal{L}_n\,g_{\mu \nu})_{|H}$. Because of how the vectors at the horizon have been chosen, the radial components of the metric remain $g_{\rho v}= n _{\mu} \ell^{\mu}= 1, \,  g_{\rho \rho}= n _{\mu} n^{\mu} =0,  g_{\rho A}=n_{\mu} \delta_{A}^{\mu} =0$, with $A = x, \phi, \tilde{\psi}$. This guarantees that the boundary conditions (\ref{Desarrollo})-(\ref{Desarrollo2}) are satisfied. The components $g_{\mu\nu }^{(0)}$ can be directly obtained from (\ref{eq:S1_metrica_inducida}), while the next-to-leading orders $g_{\mu\nu}^{(1)}$ are given by
\begin{equation}
    \begin{aligned}
        g_{v v}^{(1)} & = - \frac{\Delta'_H}{\sqrt{\Sigma_H}}, \\
        g_{v \tilde \psi}^{(1)} & = \frac{\sqrt{\Sigma_H}}{\Lambda_H} \frac{C^2 (1 + y_H)^2}{F\of{y_H} F\of{x}} \partial_y \omega\of{x, y}_{|H}, \\
        g_{v x}^{(1)} & = -\frac{\partial_x\,(\Lambda_H)}{\Lambda_H}, \\
        g_{x x}^{(1)} & = -\frac{\sqrt{\Sigma_H}}{\Lambda_H \, G\of x} \partial_y \Lambda\of{x, y} _{|H} , \\
        g_{\phi \phi}^{(1)} & = -\frac{\sqrt{\Sigma_H}}{\Lambda_H} \frac{G\of x}{F\of x} \partial_y \Lambda\of{x, y} _{|H}, \\
        g_{x \tilde \psi}^{(1)} & = - \frac{\sqrt{\Sigma_H}}{\Lambda_H} \frac{C^2 \, (1 + y_H)^2}{F\of{y_H} F\of{x}} \, \partial_y h\of{x, y} _{|H}\, , \\
        g_{\tilde \psi \tilde \psi}^{(1)} & = -\frac{\sqrt{\Sigma_{H}}}{R^2 \Lambda_{H}} \partial_y {g_{\psi \psi}}_{|H} \, .
    \end{aligned}
\end{equation}

From the results above, we can read the surface gravity on the horizon ${\kappa = -\frac{1}{2} g_{vv}^{(1)}}$, while the Jacobian $\text{det}{g_{\text{AB}}^{(0)}}$ is given in (\ref{eq:S1_metrica_inducida}). Then, we are ready to calculate the Noether charge $Q[P=1,0]$, associated to the rigid translation $\partial_v$ on the horizon. This yields the product of the temperature and the entropy of the black ring horizon. We get
\begin{equation}
S = \frac{2 \pi^2 R^3}{G} \frac{\nu ^{3/2} \sqrt{\lambda  \left(1-\lambda ^2\right)}}{(1-\nu )^2 (1 + \nu)}\,,
\end{equation}
which exactly reproduces the entropy of the black ring \cite{Emparan:2001wk}.

Analogously, we can compute the charge associated to the local Killing vector $\partial_{\tilde{\psi}}$; namely 
\begin{equation}
Q[0,\delta^A_{\tilde{\psi }}] = -\frac{\pi \nu (1 + \nu)^{3/2} R^3}{G \sqrt{2}(1 - \nu)^{3/2} \left(\nu
   ^2+1\right)}\,.
\end{equation}
However, before we rush to conclude that this quantity is the angular momentum of the black ring, we must notice that coordinate $\tilde{\psi}$ has not the right periodicity as seen from infinity. Therefore, in order to identify the correct normalization of the Killing vector we must redefine it as follows $\partial_{\tilde{\psi}} = -\frac{\delta\phi }{2\pi}\,\partial_{\psi }$. After having done so, charge (\ref{eq_cargas_Barnich}) becomes 
\begin{equation}
Q[0,\delta^A_{\psi}]=-\frac{1}{\sqrt{1+\nu ^2}}\, Q[0,\delta^A_{\tilde{\psi}}]\, ;
\end{equation}
more precisely, we find
\begin{equation}\label{LAJ}
	J_{\psi}=Q[0,\delta^{A}_{\psi }]= \frac{\pi R^3}{2 G} \frac{\sqrt{\lambda(\lambda - \nu)( \lambda + 1)}}{(1 - \nu)^2}\,.
\end{equation}
This can be rewritten in terms of only two parameters by replacing
\begin{equation}
\lambda = \frac{2\nu}{1+\nu^2}\, , \ \ \ \
\lambda-\nu =  \frac{\nu (1-\nu^2)}{1+\nu^2}\, , \ \ \ \
\lambda + 1 = \frac{(1+\nu )^2}{1+\nu^2}  \, . 
\end{equation}
Expression (\ref{LAJ}) exactly matches the result of \cite{Emparan:2006mm}. This shows that the near horizon computation of the conserved charges does reproduce the correct quantities for the black ring, at least in the case where a single angular momentum is turned on. In the next section we will consider the case in which there is also rotation in the $S^2$ section.

\subsection{Rotation in the $S^2$ section}

The metric for the black ring spinning along $\phi$ can be written as follows \cite{Figueras:2005zp}
\begin{align}
        ds^2 & = \frac{R^2 H\of{y, x}}{(x - y)^2} \bigg[- \frac{dy^2}{(1 - y^2) N\of{y}} - \frac{(1 - y^2) N\of{x}}{H\of{x, y}} d\psi^2 + \frac{dx^2}{(1 - x^2) N\of{x}} + \frac{(1 - x^2) N\of{y}}{H\of{y, x}} d\phi^2 \bigg ] \nonumber \\
        & - \frac{H\of{y, x}}{H\of{x, y}} \off{dt - \frac{\lambda \, a \, y \, (1 - x^2)}{H\of{y, x}} d\phi}^2
\end{align}
where
\begin{equation}
    N\of{\xi} = 1 + \lambda \xi + \of{\frac{a \xi}{R}}^2,
    \quad
    H\of{\xi_1, \xi_2} = 1 + \lambda \xi_1 + \of{\frac{a \xi_1 \xi_2}{R}}^2.
\end{equation}
The coordinates are the same as in the previous case, and the parameters now satisfy the relation
\begin{equation}
    \frac{2 a}{R} < \lambda < 1 + \frac{2a^2}{R^2}.
\end{equation}
which guarantees the absence of closed timelike curves and the existence of event horizons. The horizons are located at the roots of $N\of y$, which are
\begin{equation}
    y_\pm = \frac{- \lambda \pm \sqrt{\lambda^2 - 4 a^2 / R^2}}{2 a^2/ R^2}.
\end{equation}
In order to avoid conical singularities at $x = - 1$ and $y = -1$, we need to choose the periods
\begin{equation}
    \delta \phi = \delta \psi = \frac{2 \pi}{\sqrt{1 - \lambda + a^2/R^2}} \, .
\end{equation}
With these periods, the black ring is asymptotically globally flat. As before, we can rewrite the metric as follows
\begin{equation}
    \begin{aligned}
        ds^2 = \Lambda\of{x, y} \, R^2 \off{\frac{\Delta\of y}{\Sigma\of{x, y}} dt^2 - \frac{dy^2}{\Delta\of y} - \frac{M\of x}{W\of{x, y}} d\psi^2 + \frac{dx^2}{M\of x}} + g_{\phi \phi} \of{d\phi - \omega\of{x, y} \, dt}^2
    \end{aligned}
\end{equation}
where now the functions are
\begin{equation}
    \begin{aligned}
        \Delta\of y & = N\of y (1 - y^2), \ \ \ \ 
        M\of x  = N\of x (1 - x^2), \\
        W\of{x, y} & = \frac{H\of{x, y}(1 - x^2)}{(1 - y^2)}, \ \ \ \
        \Lambda\of{x, y}  = \frac{H\of{x, y}}{(x - y)^2}, \\
         \Sigma\of{x, y} & = - \lambda^2 R^2 (1 - y^2) - R^2 \Delta\of y \frac{H\of{x, y}}{H\of{y, x}} \of{\Lambda\of{x, y} - \frac{\lambda^2}{H\of{x, y}}}.
    \end{aligned}
\end{equation}
Performing the change of coordinates \eqref{eq:coordinate_change}-\eqref{eq:differential_coordinate_change}
to obtain an expression similar to \eqref{eq:black_ring_rotating_circle:metric:last_coordinate_change}, with
\begin{equation}
    \begin{aligned}
        g_{v v}^{(1)} & = - \frac{\Delta'_H}{\sqrt{\Sigma_H}} \, , \\
        g_{x x}^{(1)} & = - \frac{\sqrt{\Sigma_H}}{\Lambda_H} \frac{1}{M\of x} \partial_y \Lambda\of{x, y} _{|H} \, , \\
        g_{\tilde \phi \tilde \phi}^{(1)} & = - \frac{\sqrt{\Sigma_H}}{R^2 \Lambda_H} \partial_y g_{\phi \phi \,|H} \, , \\
        g_{v \tilde \phi}^{(1)} & = \frac{\sqrt{\Sigma_H}}{R^2 \Lambda_H} g_{\phi \phi} \partial_y \omega  _{|H} \, , \\
        g_{v x}^{(1)} & = \Lambda_H \partial_x\of{\frac{1}{\Lambda_H}} \, , \\
        g_{x \tilde \phi}^{(1)} & = \frac{\sqrt{\Sigma_H}}{R^2 \Lambda_H} g_{\phi \phi} \, \partial_y h\of{x, y}  _{|H} \, , \\
        g_{\psi \psi}^{(1)} & = \frac{\sqrt{\Sigma_H}}{\Lambda_H} M(x) \partial_y \of{\frac{\Lambda}{W}}_{|H} \, .
    \end{aligned}
\end{equation}
This enables us to compute the charges 
\begin{equation}
	Q[P, L^A] = \frac{1}{16 \pi G} \int dx \,d \tilde \phi \,d\psi\, \sqrt{\det{g}_{AB}^{(0)}}\, \Big( 2 \kappa P - L^A g^{(1)}_{vA} \Big)
  \label{eq_cargas_Barnich_2}
\end{equation}
where
\begin{equation}
	\det {g}^{(0)}_{A B} = - \frac{\lambda^2 R^6 (1 - y_H^2)}{(x - y_H)^4}.
\end{equation}
The black ring entropy would follow from the computation of the charge $Q[1, 0] = TS$; namely
\begin{equation}\label{entrompa}
	S = \frac{2 \pi R^3 \lambda}{G(1 - \lambda + a^2/R^2)} \frac{1}{\sqrt{y_H^2 - 1}},
\end{equation}
with the temperature,
\begin{equation}
	T = \frac{\sqrt{(\lambda^2 - 4 a^2 /R^2)(y_H^2 - 1)}}{4 \pi R \, \lambda}\, ,
\end{equation}
being given by the surface gravity. Formula (\ref{entrompa}) matches the correct results for the entropy as it satisfies the area law $S=\text{A}/(4G)$.

Last, by considering the correct normalization for the Killing vector $\partial_{\phi }$, we obtain
\begin{equation}
    	J_{\phi} = Q[0,\delta ^A_{\phi }]=-  \frac{\pi R^2 \lambda \, a}{G(1 - \lambda + a^2/R^2)^{3/2}},
\end{equation}
which is also found to match the correct result for the black ring angular momentum.

\subsection{Static black ring and defects}

As the last example, we can consider the particular case $\lambda = \nu$ in \eqref{eq:black_ring_rotating_circle:metric:last_coordinate_change}, which corresponds to the static black ring geometry \cite{Emparan:2004wy}. In this case, the metric functions take the form
\begin{equation}
    g_{\psi \psi} = \Sigma\of{x, y} = \frac{(y^2 - 1) F\of x^2}{(x - y)^2} R^2, \quad \Lambda\of{x, y} = \frac{F\of x}{(x - y)^2}\,.
\end{equation}
Performing a similar analysis as the one in the previous section, we find the following constraints for the periods of the angular coordinates: for $y = -1,\, x = -1$ we have 
\begin{equation}
    \delta \psi  = \delta \phi  = \frac{2 \pi}{\sqrt{1 - \nu}}\,,\label{La444}
\end{equation}
while for $x=+1$ we have
\begin{equation}
\delta \phi  = \frac{2 \pi}{\sqrt{1 + \nu}}\,,\label{La445}
\end{equation}
which cannot be simultaneously satisfied if $\nu \neq 0$. That is to say, the horizon unavoidably exhibits a singularity, which represents a defect: While satisfying the condition in (\ref{La444}) would correspond to a black ring sitting on the rim of a disc shaped deficit membrane, satisfying the condition in (\ref{La445}) would correspond to a black ring sitting on the rim of a disc shaped hole in an infinity extended deficit membrane. These defects can be thought of as dimensional extensions of the cosmic string -- or rod -- that provides the acceleration of the 4-dimensional accelerated black hole. Let us be reminded of the fact that the black ring can be pictured as the extension of the $C$-metric. This observation is important for our purpose as the analysis of the near horizon charges for the $C$-metric has been done in \cite{Anabalon:2021wjy}. By considering the defect location such that the solution is asymptotically flat, we can follow the near horizon analysis of the previous sections and compute the entropy of the static black ring. This yields
\begin{equation}
    S  = \frac{\text{A}}{4G}\ \ \ \ \text{with}\ \ \ \
    T  = \frac{\sqrt{1 - \nu^2}}{4 \pi R \nu}\,.
\end{equation}
These results are in exact agreement with those obtained in \cite{Figueras:2005zp}.

\section{Conclusions}

The results obtained in this paper show that:

$a)$ The 5-dimensional black ring solution can be accommodated in a coordinate system that fulfills the asymptotic boundary conditions defined in \cite{Donnay:2015abr}, which implies that the black ring geometry exhibits an infinite isometry enhancement in its near horizon limit. 

$b)$ The zero modes of the Noether charges associated to the aforementioned infinite-dimensional symmetry correctly reproduce the conserved quantities and thermodynamic variables of the black ring. This has been shown explicitly in the case of angular momentum in either the $S^1$ or the $S^2$ section of the horizon. The computation also works in the case of static configurations, which exhibit defects in the horizons.

$c)$ The approach we followed is systematic enough to be straightforwardly extended to higher dimensions or adapted to other cases of the vast bestiary of topologies, such as the black saturn, blackfolds, and their generalizations.

Having a method to calculate the conserved charges of solutions such as the black ring or its generalizations from the point of view of the horizon is interesting for many reasons. Firstly, it is important to remember that many techniques for generating solutions to Einstein solutions with horizons of non-trivial topology resort to near-horizon expansions that then extends outwards into the exterior region, cf.  \cite{Emparan:2007wm, Armas:2010hz, Armas:2015kra, Armas:2015nea}. Secondly, many of these solutions have multiple asymptotic regions \cite{Armas:2010pw}, so the way to carry out the standard calculation of the charges is, in many cases, not totally clear. 

Lastly, let us make a remark about the Gauss phenomenon that is behind the agreement between the calculation of conserved charges from the point of view of the horizon and the standard calculation in the asymptotic region. A priori, this might look surprising as the topology of the horizon and that of spatial infinity are generically different. However, there is a theorem that states that both computations have to agree in the case of a symmetry that is generated by an exact Killing vector. This result is general, independent of the topology: It can actually be proven \cite{Barnich:1994db} that, given an exact isometry, there exists a unique, finite and conserved surface charge 1-form in field space. Upon integrating this 1-form on any co-dimension 2 surface, one computes the infinitesimal variation of the charge associated with the symmetry that is enclosed in that surface. Since the form is conserved on-shell, one can smoothly deform the surface, as long as it does not cross any source, and the result is unchanged.  In particular, one can start with the surface $S^1 \times S^2$ enclosing the cross section of the horizon of a 5-dimensional black ring and smoothly deform it into an $S^3$ that approaches spatial infinity; the result of the charge will be identical as the form is conserved. The next step would be, of course, to go from the infinitesimal variation to the actual Noether charge. That is to say, one has to integrate the functional variation of the 1-form in the phase space. This works as long as the charge is a total variation; and in the case of isolated horizons the charges are integrable \cite{Donnay:2016ejv}. This enables to obtain the finite charge of the non-linear solution \cite{Barnich:2003xg, Barnich:2007bf} and explains why the near horizon computation does work.

\smallskip
\smallskip
\smallskip

\subsection*{Acknowledgements}

The authors thank Luciano Montecchio for discussions. G.G. is grateful to Geoffrey Compere for an important discussion. He also thanks the Solvay Institutes of Belgium for the hospitality during his stay. This work has been partially supported by grants PIP-(2017)-1109, PICT-(2019)-00303, PIP-(2022)-11220210100685CO, PIP-(2022)-11220210100225CO, PICT-(2021)-GRFTI-00644.

\bibliography{references}
\bibliographystyle{unsrturl}

\end{document}

%% file: packages.tex
\usepackage[english]{babel}
\usepackage[utf8]{inputenc}

\usepackage{indentfirst} %Paquete necesario para que haya 
\usepackage[top=3cm,bottom=3cm, left=2cm, right=2cm]{geometry}

\usepackage{hyperref} %Para que las referencias sean hipervinculos
\hypersetup{
            colorlinks=true,
            linkcolor=blue,
            urlcolor=violet,
            citecolor=violet,
            }

\usepackage{subfiles}

\usepackage{amsmath} % Necesario para estructuras donde escribir ecuaciones, e.g. align
\usepackage{amsfonts}

\numberwithin{equation}{section}

\usepackage{enumitem} % Necesario para cambiar a bullets los itemize.

\usepackage{graphicx}
\usepackage{float}
\usepackage{subcaption}

\usepackage{esint} % Integrales con sentido de circulación

\usepackage{xr} % Paquete necesario para referenciar estructuras de otro capítulo cuando se compilan individualmente.

\usepackage{cite}

\usepackage{xcolor}

\newcommand{\of}[1]{ \! \left( #1 \right ) }
\newcommand{\off}[1]{ \! \left [ #1 \right ] }